\documentclass[a4paper,11pt]{article}

\usepackage{latexsym}
\usepackage{amsfonts} 
\usepackage[mathscr]{euscript}
\usepackage{amsmath,mathrsfs,bm,amssymb,color}

\title{\textsf{
Long-range charge  order in the extended  Holstein--Hubbard model
}}
\date{\empty}
\date{\empty}
\author{
Tadahiro Miyao\\ 
 {\it Department of Mathematics,}
{\it Hokkaido University,}\\
{\it Sapporo 060-0810, Japan}\\
E-mail:
 miyao@math.sci.hokudai.ac.jp
}

\newcommand{\one}{{\mathchoice {\rm 1\mskip-4mu l} {\rm 1\mskip-4mu l}
{\rm 1\mskip-4.5mu l} {\rm 1\mskip-5mu l}}}

\newcommand{\ex}{\mathrm{e}}

\newcommand{\Fock}{\mathfrak{F}}

\newcommand{\la}{\langle}
\newcommand{\ra}{\rangle}
\newcommand{\Tr}{\mathrm{Tr}}

\newcommand{\BbbR}{\mathbb{R}}
\newcommand{\BbbN}{\mathbb{N}}
\newcommand{\BbbZ}{\mathbb{Z}}

\newcommand{\BbbC}{\mathbb{C}}
\newcommand{\vepsilon}{\varepsilon}
\newcommand{\vphi}{\varphi}

\newcommand{\dm}{\mathrm{d}}

\newcommand{\no}{\nonumber \\}

\newcommand{\bphi}{{\boldsymbol \phi}}

\newcommand{\mb}{\mathbf}

\begin{document}

\newtheorem{define}{Definition}[section]
\newtheorem{Thm}[define]{Theorem}
\newtheorem{Prop}[define]{Proposition}
\newtheorem{lemm}[define]{Lemma}
\newtheorem{rem}[define]{Remark}
\newtheorem{assum}{Condition}
\newtheorem{example}{Example}
\newtheorem{coro}[define]{Corollary}

\maketitle

\begin{abstract}
This study investigated  the extended Holstein--Hubbard model at half-filling
as a model for describing 
the interplay of  electron-electron  and 
   electron-phonon couplings.
When the electron-phonon and nearest-neighbor electron-electron interactions are strong, 
we prove the existence of long-range charge order in  three or more dimensions 
 at a sufficiently low temperature,
As a result, we rigorously  justify the phase competition between the
 antiferromagnetism 
 and charge orders.
\end{abstract} 
\section{Introduction}
 Electron-phonon
coupling  plays an essential role in the electron-pairing mechanism
in  the Bardeen--Cooper--Schrieffer theory
 \cite{BCS}.
Recently,  strong electron-phonon coupling was observed  in high-$T_c$
 cuprates \cite{Nature} and strong electron-phonon interactions were
 reported in alkali-doped fullerides and  aromatic superconductors
 \cite{Capone,
Kosugi, Kubozono, Nomura, Takabayashi}.
These examples suggest that   electron-phonon coupling 
 has received much attention in the field of superconductivity.

In the presence of strong  electron-electron Coulomb and electron-phonon
interactions, correlated electron systems provide an 
attractive field of study exhibiting a 
competition among
  various phases.
Despit the extensive research regarding  the competition between these phases, 
only few exact results are  currently known.
The Holstein--Hubbard model is  a simple model that enables us the
exploration
of 
the interplay of electron-electron and electron-phonon interactions.
Our aim  is to  rigorously study
the competition between the phases in the system described by the
this  model.

Rigorous study of the Holstein model was initiated by L\"owen \cite{Lowen}.
Later, Freericks and Lieb proved that the ground state of the Holstein
model is unique and has a total spin $S=0$ \cite{FL}.
However, their studies focused on  electron-phonon interaction only
and    did not consider the  interplay between  electron-electron
 and electron-phonon interactions.
Taking this  interplay into account, 
 Miyao proved  the
  following  \cite{Miyao3}:
\begin{itemize}
    \item If the electron-phonon coupling is weak ($U_{\mathrm{eff}}-\nu V>0$), 
       there is no long-range  charge order in  the Holstein--Hubbard
	  system at half-filling.
    \item If   the electron-phonon coupling is weak, the ground state of 
the Holstein--Hubbard model is unique and exhibits  antiferromagnetism.
\end{itemize}
More   precise statements of these two principles are provided in
Section \ref{MainRe}.
The achievement of this study  is the proof   that there exists a  long-range 
 charge order
at a sufficiently low temperature
 provided that the electron-phonon interaction is strong ($U_{\mathrm{eff}}-\nu V <0$).
The obtained phase diagram is compatible with the previous results
conjectured  by 
heuristic arguments \cite{Bari, Murakami}.
To prove the main result, we apply the method of reflection positivity.

Reflection positivity originates from axiomatic quantum field theory \cite{OS}.
Glimm, Jaffe and Spencer first applied reflection positivity to the study of 
 phase transition \cite{GJS, GlimmJaffe}.  This  idea was  further developed by Dyson, Fr\"ohlich, Israel,
Lieb, Simon and Spencer in \cite{DLS, FILS, FILS2, FSS} and  applications of reflection positivity to the 
Hubbard model are  given in \cite{JJ,KuboKishi, Lieb}.
  In the present  study, we 
further develop the method used in \cite{FILS2}  to 
 apply reflection positivity 
to the Holstein--Hubbard model which is more difficult  to analyze than
the Hubbard model.  

Usually, the hopping matrix elements of the Hubbard model are real
numbers. Because of past successes in the research of the phase transitions of the Hubbard model, 
it appears that reflection positivity was   inapplicable to the case where
the hopping matrix elements are {\it complex} numbers. 
In the study of the Holstein--Hubbard model, the Lang--Firsov transformation is known 
to be very  useful. However,  this transformation  changes  the hopping matrix elements 
from real  into  complex numbers.  Therefore,
at first glance,
 it appears that reflection
positivity is unsuitable for the study of  phase transitions of 
the Holstein--Hubbard model.
On the other hand, in a series of papers \cite{Miyao3, Miyao4}, Miyao
has  shown that 
reflection positivity is  still applicable to
several models with  complex  hopping matrix elements (also see
\cite{Miyao2}).\footnote{Namely, he applied  the {\it spin} reflection
positivity to the Holstein--Hubbard and Su--Schrieffer--Heeger models and
investigated their ground state properties.}
In the present  paper, we further extend   this idea
 and adapt reflection positivity to a rigorous  analysis of the phase
 transitions of 
the  Holstein--Hubbard model. 
 
 Note  that an application of  reflection positivity to the 
Hubbard model with complex hopping matrix elements was  first discussed
by Lieb \cite{Lieb2} in his solution of  the flux-phase conjecture (also see \cite{GMP,
LiebLoss, MN,  Miyao4}).
 Our  present paper  aims to apply    reflection positivity 
to the study of   phase transitions of   a model of interacting
electrons  with complex
hopping matrix elements.

The rest  of the paper is organized as follows:
In Section \ref{MainRe}, we define the Holstein--Hubbard model and state the main results.
We also  compare the obtained  results with those of  previous studies as well.
Section \ref{ProofM} is devoted to the proof of the main theorem.
In Appendix \ref{PfHalf}, we  show that our system is half-filled
with electrons;
in Appendix \ref{AppA}, we give an extension of the Dyson--Lieb--Simon
inequality; and 
in Appendix \ref{AppB}, we prove a useful inequality.
Appendix \ref{AntiU} is devoted to construct an antiunitary transformation
which plays an important role in Section \ref{ProofM}.

\begin{flushleft}
{\bf Acknowledgments.}
This work was partially supported by KAKENHI (20554421) and   KAKENHI(16H03942).
I would be grateful to the anonymous referees for useful comments.
\end{flushleft} 

\section{Main results}\label{MainRe}
\setcounter{equation}{0}

Let $\Lambda=[-L, L)^{\nu}\cap \BbbZ^{\nu}$. 
The extended Holstein--Hubbard model on $\Lambda$ is given by 
\begin{align}
H_{\Lambda}=&\sum_{\la x; y\ra}\sum_{\sigma=\uparrow, \downarrow}(-t)(c_{x\sigma}^*
 c_{y\sigma}
+c_{y\sigma}^*c_{x\sigma}
)\no
&+U \sum_{x\in \Lambda}(n_x-\one)^2
+V\sum_{\la x; y\ra} (n_x-\one)(n_y-\one)\no
&+ g \sum_{x\in \Lambda} (n_x-\one)(b_x+b_x^*)
+\omega \sum_{x\in \Lambda} b_x^* b_x,
\end{align} 
where $n_x=n_{x\uparrow}+n_{x\downarrow}$ with
$n_{x\sigma}=c_{x\sigma}^* c_{x\sigma}$.
Here, $\la x;  y\ra$ refers to  a sum over nearest-neighbor pairs.
We impose periodic boundary conditions, so $L\equiv -L$. 
$H_{\Lambda}$ acts in the Hilbert space according to 
\begin{align}
\mathfrak{H}=\mathfrak{F}\otimes \mathfrak{P}.
\end{align} 
The electrons exist in the fermionic Fock space $\Fock$  given by 
$
\Fock:= \Fock_{\mathrm{as}}(\ell_{\uparrow}^2(\Lambda)\oplus
\ell^{2}_{\downarrow}(\Lambda)):=\bigoplus_{n\ge 0}
\wedge^n (\ell_{\uparrow}^2(\Lambda)\oplus
\ell^{2}_{\downarrow}(\Lambda))
$, where
$\ell_{\uparrow}^2(\Lambda)=\ell^2_{\downarrow}(\Lambda)=\ell^2(\Lambda)$,
and $\wedge^n$ is the $n$-fold antisymmetric tensor product.
The phonons exist  in the bosonic Fock space $\mathfrak{P}$ defined by 
$
\mathfrak{P}=\bigoplus_{n\ge 0} \otimes_{\mathrm{s}}^n \ell^2(\Lambda)
$,
where $\otimes_{\mathrm{s}}^n$ is the $n$-fold symmetric tensor product, 
$c_{x\sigma}$ is the electron annihilation operator, and  $b_x$ is the phonon
annihilation operator.
 These operators satisfy the following relations:
\begin{align}
\{c_{x\sigma}, c_{x'\sigma'}^*\}=\delta_{\sigma\sigma'}\delta_{xx'},\ \ \ 
[b_x, b_{x'}^*]=\delta_{xx'}.
\end{align} 
$t$ is the hopping matrix element, and $g$ is the strength of the
electron-phonon interaction. The  on-site and  nearest-neighbor repulsions
are denoted by $U$ and $V$, respectively.
The phonons are assumed to be dispersionless with energy $\omega$.
Henceforth, we  assume the following:
\begin{itemize}
\item $
g\in \BbbR,\ \ t>0,\ \ U>0,\ \ V>0,\ \ \omega>0.
$
\item  $L$ is an odd number.
\end{itemize} 

The thermal expectation value  is defined  by
\begin{align}
\la A\ra_{\beta, \Lambda}=
\Tr\big[
A\, e^{-\beta H_{\Lambda}}
\big]\Big/ Z_{\beta, \Lambda},
\ \ 
Z_{\beta, \Lambda}=\Tr\big[e^{-\beta H_{\Lambda}}\big].
\end{align}  
We restrict ourselves to the case of half-filling.
 In fact,  we show that  
 \begin{align}
 \la n_x\ra_{\beta, \Lambda}=1\label{Half}
  \end{align}
  in Appendix \ref{PfHalf}. 
We let $q_x=n_x-\one$ 
and  define the two-point correlation function as\footnote{In this paper, we simply assume that the right hand side of (\ref{Limit}) exists.
Alternatively, we choose a subsequence such that the right hand side of  (\ref{Limit})
exists.
} 
\begin{align}
\la q_x q_o\ra_{\beta}=\lim_{L\to \infty} \la q_x q_o\ra_{\beta, \Lambda}.\label{Limit}
\end{align} 
The effective interaction strength is defined as 
\begin{align}
U_{\mathrm{eff}}=U-\frac{2g^2}{\omega}.
\end{align} 
In \cite{Miyao3}, the following theorem is proven provided that  $\nu V-U_{\mathrm{eff}}<0$.
\begin{Thm}{\rm \cite{Miyao3}}
Suppose that $
\nu V-U_{\mathrm{eff}}<0
$. Then  the following is obtained:
\begin{itemize}
\item[{\rm (i)}]  
For all $\beta \ge 0$, we have
\begin{align}
\lim_{\|x\|\to \infty} \la q_x q_o\ra_{\beta}=0.
\end{align}  
Hence, there is no  long-range charge  order.
\item[{\rm (ii)}] Let $\mathfrak{H}_{ M}$ be the $M$-subspace\footnote{
To be precise, $\mathfrak{H}_M$ is defined by 
\begin{align}
\mathfrak{H}_M=\{\psi\in \mathfrak{H}\, |\, N\psi=|\Lambda|\psi,\ S^3\psi=M\psi\},
\end{align} 
where $N=N_{\uparrow}+N_{\downarrow}$ and
	     $S^3=\frac{1}{2}(N_{\uparrow}-N_{\downarrow})$ with
	     $N_{\sigma}=\sum_{x\in \Lambda} n_{x\sigma}$.
The condition $N\psi=|\Lambda|\psi$  indicates that  we consider
	     the case of  half-filling.
}
	     and 
let $H_{\Lambda, M}=H_{\Lambda}\restriction \mathfrak{H}_M$, the restriction of $H_{\Lambda}$ to
	     $\mathfrak{H}_M$. The ground state of $H_{\Lambda, M}$ is unique
	     for all  possible values of $M$.
\item[{\rm (iii)}] Let 
\begin{align}
S_x^+=c_{x\uparrow}^* c_{x\downarrow},\ \ S_{x}^-=c_{x\downarrow}^* c_{x\uparrow}.
\end{align} 
Let $\vphi_M$ be the ground state of $H_{\Lambda, M}$. We obtain 
\begin{align}
(-1)^{\|x\|}\la \vphi_M|S_x^+S_o^-\vphi_M\ra>0
\end{align} 
for all $x\in \Lambda$, where $\|x\|=\sum_{j=1}^{\nu}|x_j|$.
This  means  that  the ground state is  antiferromagnetic. 
\end{itemize} 
\end{Thm}

It is logical and  important to study the case where  $\nu V-U_{\mathrm{eff}}>0$.  
Our main result in this paper is the following:
\begin{Thm}\label{Main}
Assume that 
$
\nu V-U_{\mathrm{eff}}>0.
$
For each $\nu \ge 3$,  we have
\begin{align}
&\liminf_{\|x\|\to \infty} (-1)^{\|x\|}\la q_x q_o\ra_{\beta}\\
&\ge
1-
\beta^{-1} (\nu V-U_{\mathrm{eff}})^{-1} \ln 4(1-e^{-\beta \omega })^{-1}
-8\nu t (\nu V-U_{\mathrm{eff}})^{-1}
-\gamma_1 \int_{\Bbb{T}^{\nu}} dp E(p)^{-1}
-\gamma_2, \label{LowerBound}
\end{align} 
where $\mathbb{T}=(-\pi, \pi),\ E(p)=\sum_{j=1}^{\nu}(1-\cos p_j)$  and 
\begin{align}
\gamma_1=(2\pi)^{-\nu }\frac{1}{2}\Big\{
(\beta V)^{-1}+\Big(
\frac{t}{V}
\Big)^{1/2}
\Big\},\ \ \ 
\gamma_2= \frac{1}{4} \Big(
\frac{t}{V}
\Big)^{1/2}.
\end{align} 
\end{Thm}

\begin{coro}
Let $\nu \ge 3$. Assume that 
$\nu V-U_{\mathrm{eff}}>0$. If $\beta, V, g$ are sufficiently large such
 that  the right-hand side of (\ref{LowerBound}) is strictly positive,  
then we obtain
\begin{align} 
\liminf_{\|x\|\to \infty} (-1)^{\|x\|} \la q_x q_o\ra_{\beta}>0.
\end{align} 
Thus,  a staggered   long-range charge  order exists.

\end{coro}

\section{Proof of Theorem \ref{Main}}\label{ProofM}
\setcounter{equation}{0}
\subsection{ Lang--Firsov transformation}
We set $H_{\Lambda}=T+P+I+K$, where
\begin{align}
T&=\sum_{\la x; y\ra}\sum_{\sigma=\uparrow, \downarrow}(-t)\big(
c_{x\sigma}^*
 c_{y\sigma}
+c_{y\sigma}^*c_{x\sigma}
\big),\label{EKinetic}\\
P&=U \sum_{x\in \Lambda}q_x^2
+V\sum_{\la x; y\ra} q_xq_y,\\
I&=g \sum_{x\in \Lambda} q_x(b_x+b_x^*),\\
K&= \omega \sum_{x\in \Lambda} b_x^* b_x\label{BKinetic}.
\end{align}
For each $x\in \Lambda$, let
\begin{align}
 \phi_x=\sqrt{\frac{1}{2\omega}}(b_x^*+b_x),\ \ \pi_x=i
 \sqrt{\frac{\omega}{2}}(b_x^*-b_x).
\end{align} 
Both $\phi_x$ and $\pi_x$ are essentially self-adjoint and  we  denote their
closures
 by the same symbols. Next, let
\begin{align}
L=-i \omega^{-3/2} g\sum_{x\in \Lambda} q_x \pi_x.
\end{align} 
$L$ is essentially antiself-adjoint. We also denote its closure by the
same symbol.
The {\it Lang-Firosov transformation } is a unitary operator defined by
\begin{align}
\mathscr{U}=e^{-i \pi N_{\mathrm{p}}/2}e^L,
\end{align}   
where $N_{\mathrm{p}}=\sum_{x\in\Lambda} b_x^*b_x$ \cite{LF}. We can check the
following:
\begin{align}
\mathscr{U} c_{x\sigma} \mathscr{U}^{-1}&=e^{i\alpha \phi_x} c_{x\sigma},
 \ \ \alpha=\sqrt{2} \omega^{-3/2} g,\\
\mathscr{U} b_x\mathscr{U}^{-1}&=b_x-\frac{g}{\omega}q_x.
\end{align} 
Using  these formulas, we obtain the following:
\begin{lemm}
Let $H'_{\Lambda}=\mathscr{U} H_{\Lambda}\mathscr{U}^{-1}$. We have 
\begin{align}
H'_{\Lambda}=T'+P'+K,
\end{align}
where 
\begin{align}
T'&= \sum_{\la x; y\ra} \sum_{\sigma=\uparrow, \downarrow}(-t)
\Big(e^{-i \alpha(\phi_x-\phi_y)}
c_{x\sigma}^* c_{y\sigma}
+
e^{+i \alpha(\phi_x-\phi_y)}
c_{y\sigma}^* c_{x\sigma}
\Big),\\
P'&=U_{\mathrm{eff}} \sum_{x\in \Lambda} q_x^2+V\sum_{\la x; y\ra}
 q_xq_y,\\
K&= \frac{1}{2}\sum_{x\in \Lambda} (\pi_x^2+\omega^2 \phi_x^2).
\end{align} 
\end{lemm} 

\subsection{The Schr\"odinger representation}
The bosonic Fock space can be identified as
\begin{align}
\mathfrak{P}=L^2(\mathcal{Q}_{\Lambda}, d\mu_{\Lambda}),
\end{align} 
where $\mathcal{Q}_{\Lambda}=\BbbR^{\Lambda}$ and
$d\mu_{\Lambda}=\prod_{x\in \Lambda} d\phi_x$ is the
$|\Lambda|$-dimensional Lebesgue measure.
Moreover, each $\phi_x$ can be regarded as a multiplication operator by
the {\it real}-valued function,  and $\pi_x$ can be regarded as a
partial differential operator $-i \frac{\partial}{\partial \phi_x}$.
This representation of the canonical commutation relations is called the
{\it Schr\"odinger representation}.
In the following section, we will focus on   this representation.

\subsection{The zigzag transformation}
Following \cite{FILS2}, we introduce the {\it zigzag  transformation} as follows:
Let 
\begin{align}
v_{x\sigma}=\Bigg[
\prod_{z\neq x} (-1)^{n_{z\sigma}} 
\Bigg](c_{x\sigma}^*+c_{x\sigma}).
\end{align} 
Note  that $v_{x\sigma}^{-1}=v_{x\sigma}$.
It is not hard to check that 
\begin{align}
v_{x\sigma} c_{x'\sigma'} v_{x\sigma}^{-1} =\begin{cases}
c_{x\sigma}^ * & \mbox{if $(x, \sigma)=(x', \sigma')$}\\
c_{x'\sigma'} & \mbox{if  $(x, \sigma)\neq (x', \sigma')$ }
\end{cases}.
\end{align}
Let $\Lambda_{\mathrm{e}}=\{x\in \Lambda\, |\, \mbox{$\|x\|$  is
even}\}$ and let $\Lambda_{\mathrm{o}}=\{x\in \Lambda\, |\,
\mbox{$\|x\|$ is odd}\}$.
Now,  we set
\begin{align}
\mathscr{V}= \prod_{x\in \Lambda_{\mathrm{o}}} v_{x\uparrow} v_{x\downarrow}.
\end{align} 
We observe that 
\begin{align}
\mathscr{V} c_{x\sigma} \mathscr{V}^{-1}
=\begin{cases}
c_{x\sigma}^* & \mbox{if $x\in \Lambda_{\mathrm{o}}$}\\
c_{x\sigma} & \mbox{if $x\in \Lambda_{\mathrm{e}}$}
\end{cases} ,
\ \ \ \mathscr{V} q_x \mathscr{V}^{-1} =(-1)^{\|x\|} q_x. \label{Zigzag}
\end{align} 

\begin{lemm}
Let $H''_{\Lambda}=\mathscr{V} H'_{\Lambda} \mathscr{V}^{-1}$. We have 
$H''_{\Lambda}=T''+P''+K$, where
\begin{align}
T''&= 
\sum_{x\in \Lambda_{\mathrm{e}}}\sum_{\sigma=\uparrow, \downarrow} \sum_{j=1}^{\nu} \sum_{\vepsilon=\pm}
(-t) \Big(
e^{-i \alpha(\phi_x-\phi_{x+\vepsilon \delta_j})}
c_{x\sigma}^* c_{x+\vepsilon \delta_j\sigma}^*
+
\mathrm{h. c. }
\Big), \label{PP}\\
P''&= U_{\mathrm{eff}}\sum_{x\in \Lambda} q_x^2-V\sum_{\la x; y\ra} q_x q_y.\label{PP2}
\end{align} 
Here, $\delta_j\, (j=1, \dots, \nu)$ is the unit vector in $\BbbZ^{\nu}$ defined by 
$
\delta_j=(0, \dots, 0, \underbrace{1}_{j\mathrm{-th}},0,\dots, 0)
$. 
\end{lemm} 
{\it Proof.} $T'$ can be expressed as 
\begin{align}
T'
=\sum_{x\in \Lambda_{\mathrm{e}}}\sum_{\sigma= \uparrow, \downarrow} \sum_{j=1}^{\nu}
 \sum_{\vepsilon=\pm 1}(-t)
\Big(
e^{-i \alpha (\phi_x-\phi_{x+\vepsilon\delta_j})}c_{x\sigma}^*
 c_{x+\vepsilon \delta_j\sigma}+\mathrm{h.c.}
\Big).
\end{align} 
Thus, by using (\ref{Zigzag}), we obtain (\ref{PP}),  and  similarly, 
(\ref{PP2}). $\Box$
\medskip\\

To show the main theorem, we introduce the following modified
Hamiltonian:
\begin{define}
{\rm 
For each $\mb{h}=\{h_x\}_{x\in \Lambda}\in \BbbR^{\Lambda}$, we set
\begin{align}
P''(\mb{h})=(U_{\mathrm{eff}}-\nu V)\sum_{x\in \Lambda} q_x^2+\frac{V}{2}\sum_{\la x; y\ra} (q_x-h_x- q_y+h_y)^2
\end{align} 
and 
\begin{align}
H''_{\Lambda}(\mb{h})=T''+P''(\mb{h})+K.
\end{align} 
Trivially, we have $H''_{\Lambda}=H''_{\Lambda}(\mb{0})$. $\diamondsuit$
}
\end{define} 

\subsection{Reflection positivity}
\subsubsection{Overview}
The hopping matrix elements  in (\ref{PP}) are  complex.
In general, it is impossible to apply  reflection
positivity (RP) to a  fermionic system with complex hopping matrix elements.
However, a suitably modified  RP can be still  applicable to $H_{\Lambda}''$   because 
these complex phase factors (i.e., $
e^{-i\alpha(\phi_x-\phi_{x+\vepsilon \delta_j})}
$ ) are not random, but rather exhibit  a regular structure.

Next, we briefly explain  the modified RP.
Let $\mathfrak{X}_L$ and $\mathfrak{X}_R$ be complex Hilbert spaces,
and let $\vartheta$ be an {\it antiunitary transformation}
from $\mathfrak{X}_L$ onto $\mathfrak{X}_R$. In Appendix \ref{AppA},
we prove the following:
\begin{align}
\Tr_{\mathfrak{X}_L\otimes \mathfrak{X}_R}\Big[
A\otimes \vartheta A\vartheta^{-1}
\Big]=
\big|\Tr_{\mathfrak{X}_L}[A]\big|^2\ge 0.
\end{align} 
This is the  basic idea of the modified RP.
Thus, our problem is reduced to constructing  a suitable $\vartheta$.
This formalism  allows us to apply  RP to $H_{\Lambda}''$.

In Proposition \ref{AntiUni}  and Appendix \ref{AntiU}, we actually construct a suitable
$\vartheta$.  Moreover, in Lemmas \ref{A1} and \ref{A2}, we prove that the extended
RP
can be applicable to our  model. 
In these arguments, we  carefully use the regular structure of the phase
factors
and   the assumption that $L$ is odd.

In the original paper \cite[Section 3]{FILS}, the authors give several
examples of how we construct RP.
Our formalism is different from these examples and  more convenient
 for studying  the Holstein--Hubbard model.

\subsubsection{Preliminaries}
We divide $\Lambda$ as $\Lambda=\Lambda_{L}\cup \Lambda_R $, where 
\begin{align}
\Lambda_L=\{x=(x_1, \dots, x_{\nu})\in \Lambda\, |\, x_1 <0\},
\ \ \Lambda_R=\{x=(x_1, \dots, x_{\nu})\in \Lambda\, |\, x_1\ge 0\}.
\end{align} 
Corresponding to this, we also divide  $\ell^2(\Lambda)$ as 
\begin{align}
\ell^2(\Lambda)=\ell^2(\Lambda_L)\oplus \ell^2(\Lambda_R).
\end{align} 
Hence, we have the  following identifications:
\begin{align}
\Fock=\Fock_L\otimes \Fock_R,
\end{align} 
where 
$
 \Fock_L=\Fock_{\mathrm{as}}(\ell_{\uparrow}^2(\Lambda_L)\oplus
 \ell^2_{\downarrow}(\Lambda_L))
$ and $
\Fock_R=\Fock_{\mathrm{as}}(\ell_{\uparrow}^2(\Lambda_R)\oplus
 \ell^2_{\downarrow}(\Lambda_R)),
$
and 
\begin{align}
\mathfrak{P}=\mathfrak{P}_L\otimes \mathfrak{P}_R,
\end{align} 
where
$
 \mathfrak{P}_L=\Fock_{\mathrm{s}}(\ell^2(\Lambda_L))
=L^2(\mathcal{Q}_{\Lambda_L}, d\mu_{\Lambda_L})
$
and 
$
\mathfrak{P}_R=\Fock_{\mathrm{s}}(\ell^2(\Lambda_R))
=
L^2(\mathcal{Q}_{\Lambda_R}, d\mu_{\Lambda_R})
$.
Thus, the Hilbert space $\mathfrak{H}$ can be identified as follows:
\begin{align}
\mathfrak{H}=\mathfrak{H}_L\otimes \mathfrak{H}_R, \label{Ident}
\end{align} 
where 
$
\mathfrak{H}_L=\mathfrak{F}_L\otimes \mathfrak{P}_L
$ and $
 \mathfrak{H}_R=\mathfrak{F}_R\otimes \mathfrak{P}_R.
$
Under the identification (\ref{Ident}), we have the following identifications:
\begin{align}
c_{x\sigma}=
\begin{cases}
c_{x\sigma}\otimes \one & \mbox{if $x\in \Lambda_L$}\\
(-1)^{N_L} \otimes c_{x\sigma} &\mbox{if $x\in \Lambda_R$},
\end{cases} 
\end{align} 
where $N_L=\sum_{ x\in \Lambda_L} n_x$, and 
\begin{align}
\pi_x=\begin{cases}
\pi_x\otimes \one & \mbox{ if $x\in \Lambda_L$}\\
\one \otimes \pi_x& \mbox{ if $x\in \Lambda_R$}
\end{cases},\ \ \ \  
\phi_x =
\begin{cases}
\phi_x\otimes \one & \mbox{ if $x\in \Lambda_L$}\\
\one \otimes \phi_x& \mbox{ if $x\in \Lambda_R$}
\end{cases}.
\end{align} 
Using these,  we state the following lemmas:

\begin{lemm}
Under the identification (\ref{Ident}), 
we have 
$T''=T''_L\otimes \one +\one \otimes T_R''+T''_{LR}$, where 
\begin{align}
T_L''=&
\sum_{x\in \Lambda_{\mathrm{e}},\  x_1\le -2}\sum_{\sigma=\uparrow, \downarrow}
\sum_{j=1}^{\nu} \sum_{\vepsilon=\pm}'
(-t)  \Big(
e^{-i \alpha(\phi_x-\phi_{x+\vepsilon \delta_j})}
c_{x\sigma}^* c_{x+\vepsilon \delta_j\sigma}^*
+
\mathrm{h. c. }
\Big),\\
T_R'' =&\sum_{x\in \Lambda_{\mathrm{e}},\  x_1\ge 0}
\sum_{\sigma=\uparrow, \downarrow}
\sum_{j=1}^{\nu} \sum_{\vepsilon=\pm}''
(-t)  \Big(
e^{-i \alpha(\phi_x-\phi_{x+\vepsilon \delta_j})}
c_{x\sigma}^* c_{x+\vepsilon \delta_j\sigma}^*
+
\mathrm{h. c. }
\Big),\label{TRc}\\
T''_{LR}=&
\sum_{x\in \Lambda_{\mathrm{e}},\ x_1=0} \sum_{\sigma=\uparrow, \downarrow}(-t)
\Bigg\{
\Big[
e^{i\alpha \phi_{x-\delta_1}}
(-1)^{N_L} c_{x-\delta_1  \sigma}
^*\Big]
\otimes
 \Big[
e^{-i \alpha \phi_x }c_{x\sigma}^*
\Big]+\mathrm{h.c.}
\Bigg\}\no
&+ 
\sum_{x\in \Lambda_{\mathrm{e}},\ x_1=L-1} \sum_{\sigma=\uparrow, \downarrow}(-t)
\Bigg\{
\Big[
e^{i\alpha \phi_{x+\delta_1}}
(-1)^{N_L} c_{x+\delta_1  \sigma}
^*\Big]
\otimes
 \Big[
e^{-i \alpha \phi_x }c_{x\sigma}^*
\Big]+\mathrm{h.c.}
\Bigg\}. \label{TRL}
\end{align} 
Here, $\displaystyle \sum_{\vepsilon=\pm}'$ refers to a sum over pairs
 $\la x; x+\vepsilon \delta_j\ra$ such that $x, x+\vepsilon \delta_j\in
 \Lambda_L$. Similarly,   $\displaystyle \sum_{\vepsilon=\pm}''$ refers to a sum over pairs
 $\la x; x+\vepsilon \delta_j\ra$ such that $x, x+\vepsilon \delta_j\in
 \Lambda_R$.
\end{lemm}

\begin{rem}
{\rm
To obtain (\ref{TRL}), we assume that  $L$ is odd. $\diamondsuit$
}
\end{rem} 

\begin{lemm}
For each $\mb{h}=\{h_x\}_{x\in \Lambda}\in \BbbR^{\Lambda}$, we set 
$\mb{h}_L=\{h_x\}_{x\in \Lambda_L}$ and $\mb{h}_R=\{h_x\}_{x\in \Lambda_R}$.
We have $P''(\mb{h})=P''_L(\mb{h}_L)\otimes \one +\one \otimes P_R''(\mb{h}_R)+P''_{LR}(\mb{h})$, where  
\begin{align}
P''_L(\mb{h}_L)=& (U_{\mathrm{eff}}-\nu V)\sum_{x\in \Lambda_L}
 q_x^2+\frac{V}{2}\sum_{\la x; y\ra,\ x, y\in \Lambda_L} (q_x-h_x- q_y+h_y)^2,\\
P_R''(\mb{h}_R) =& 
(U_{\mathrm{eff}}-\nu V)\sum_{x\in \Lambda_R} q_x^2+\frac{V}{2}\sum_{\la
 x; y\ra,\ x, y\in \Lambda_R} (q_x-h_x- q_y+h_y)^2,\\
P''_{LR}(\mb{h}) =&
-V \sum_{x\in \Lambda_{\mathrm{e}},\ x_1=0} (q_{x-\delta_1}-h_{x-\delta_1})
\otimes (q_x-h_x)\no
&-V \sum_{x\in \Lambda_{\mathrm{e}},\ x_1=L-1} (q_{x+\delta_1}-h_{x+\delta_1})\otimes (q_x-h_x).
\end{align} 
\end{lemm}

\begin{lemm}
We have $K=K_L\otimes \one +\one \otimes K_R$, where 
\begin{align}
K_L=\frac{1}{2}\sum_{x\in \Lambda_L}(\pi_x^2+\omega^2\phi_x^2),
\ \ \ K_R=\frac{1}{2}\sum_{x\in \Lambda_R}(\pi_x^2+ \omega^2\phi_x^2). 
\end{align} 
\end{lemm}

For all $x\in \Lambda_L$,  we define
\begin{align}
a_{x\sigma}=c_{x\sigma} (-1)^{N_L}.
\end{align} 

In terms of $a_{x\sigma}$, $T_L''$ and $T''_{LR}$ can be expressed as
follows.

\begin{Prop}
We obtain the following:
\begin{align}
T_L'' =&\sum_{x\in \Lambda_{\mathrm{e}},\  x_1\le -2}
\sum_{\sigma=\uparrow, \downarrow}\sum_{j=1}^{\nu} \sum_{\vepsilon=\pm}'
(+t)  \Big(
e^{-i \alpha(\phi_x-\phi_{x+\vepsilon \delta_j})}
a_{x\sigma}^* a_{x+\vepsilon \delta_j\sigma}^*
+
\mathrm{h. c. }
\Big),\\
T''_{LR}=&
\sum_{x\in \Lambda_{\mathrm{e}},\ x_1=0} \sum_{\sigma=\uparrow, \downarrow}(-t)
\Bigg\{
\Big(
e^{i\alpha \phi_{x-\delta_1}}
a_{x-\delta_1 \sigma}^*
\Big)
\otimes
 \Big(
e^{-i \alpha \phi_x }c_{x\sigma}^*
\Big)+\mathrm{h.c.}
\Bigg\}\no
&+ 
\sum_{x\in \Lambda_{\mathrm{e}},\ x_1=L-1} \sum_{\sigma=\uparrow, \downarrow}(-t)
\Bigg\{
\Big(
e^{i\alpha \phi_{x+\delta_1}}
a_{x+\delta_1\sigma}^*
\Big)
\otimes
 \Big(
e^{-i \alpha \phi_x }c_{x\sigma}^*
\Big)+\mathrm{h.c.}
\Bigg\}.
\end{align} 

\end{Prop} 
\begin{rem}
{\rm 
Since $q_x=\sum_{\sigma=\uparrow, \downarrow} a_{x\sigma}^*a_{x\sigma}-\one$,  expressions of
 $P''_L(\mb{h}_L)$ and
 $P''_{LR}(\mb{h})$ are  unchanged if we write these in terms of
 $a_{x\sigma}$. $\diamondsuit$
}
\end{rem}

\subsubsection{Gaussian domination}

We define the {\it reflection map} $r: \Lambda_R\to \Lambda_L$ by 
\begin{align}
r(x)=(-x_1-1, x_2,\dots, x_{\nu}),\ \ x\in \Lambda_R. \label{ReflectionMap}
\end{align} 
We begin with the following proposition:
\begin{Prop}\label{AntiUni}
There exists an antiunitary transformation\footnote{
Namely, $\vartheta$ is a bijective antilinear map which satisfies
$
\la \vartheta \vphi|\vartheta\psi\ra=(\la \vphi|\psi\ra)^*
$ for all $\vphi, \psi\in \mathfrak{X}_L.$
}  $\vartheta$  from $\mathfrak{H}_L$
to $\mathfrak{H}_R$ such that 
\begin{align}
&c_{x\sigma}=\vartheta a_{r(x)\sigma} \vartheta
^{-1},\ \ \ \phi_x=\vartheta \phi_{r(x)} \vartheta^{-1},\ \ \
\pi_x=-\vartheta \pi_{r(x)}\vartheta^{-1}, \ \ 
 x\in
 \Lambda_R,\label{theta}
\\
&\vartheta \Omega_L=\Omega_R, \label{theta2}
\end{align}  
where $\Omega_L$  is the  Fock vacuum $\Omega_{\mathrm{f}}^L \otimes
\Omega_{\mathrm{b}}^L$ in $\mathfrak{H}_L$
\footnote{
In the Schr\"odinger representation,
$\Omega_{\mathrm{b}}^L=(\frac{1}{\pi})^{|\Lambda_L|/4}
e^{-\sum_{x\in \Lambda_L}\phi_x^2/2}
$. $\Omega_{\mathrm{f}}^L$ is the standard Fock vacuum in $\Fock_L$.
Note  that $\Omega_{\mathrm{b}}^L$ is a real-valued function on $\mathcal{Q}_{\Lambda_L}$.
}, and 
$\Omega_R$ can be defined in a similar manner.
\end{Prop} 
{\it Proof.} See Appendix \ref{AntiU}. $\Box$

\begin{lemm}\label{A1}
We have the following:
\begin{itemize}
\item[{\rm (i)}]
$
\displaystyle 
T_R''=\vartheta T_L'' \vartheta^{-1}
$.
\item[{\rm (ii)}]
\begin{align}
T_{LR}''=&\sum_{x\in \Lambda_{\mathrm{e}},\ x_1=0}
 \sum_{\sigma=\uparrow, \downarrow}(-t)
\Bigg\{
\Big(
e^{i\alpha \phi_{x-\delta_1}}
a_{x-\delta_1 \sigma}^*
\Big)
\otimes 
\vartheta \Big(
e^{i \alpha \phi_{x-\delta_1} }a_{x-\delta_1\sigma}^*
\Big)
\vartheta^{-1}+\mathrm{h.c.}
\Bigg\}\no
&+ 
\sum_{x\in \Lambda_{\mathrm{e}},\ x_1=L-1} \sum_{\sigma=\uparrow, \downarrow}(-t)
\Bigg\{
\Big(
e^{i\alpha \phi_{x+\delta_1}}
a_{x+\delta_1\sigma}^*
\Big)
\otimes
 \vartheta \Big(
e^{i \alpha \phi_{x+\delta_1} }a_{x+\delta_1\sigma}^*
\Big)\vartheta^{-1}
+\mathrm{h.c.}
\Bigg\}.
\end{align} 
\end{itemize} 

\end{lemm} 
{\it Proof.} While (ii) is trivial,  (i) has be addressed carefully.
First,  $T_L''$ can be expressed as 
\begin{align}
T_L''=\sum_{x\in \Lambda_{\mathrm{o}},\  x_1\le -1}
\sum_{\sigma=\uparrow, \downarrow}
\sum_{j=1}^{\nu} \sum_{\vepsilon=\pm}'
(+t)  \Big(
e^{-i \alpha(\phi_{x+\vepsilon \delta_j}-\phi_x)}
 a_{x +\vepsilon \delta_j\sigma}^* a_{x\sigma}^*
+
\mathrm{h. c. }
\Big).
\end{align} 
Hence, by (\ref{TRc}), we see  that 
\begin{align}
T''_R &=
\vartheta \sum_{x\in \Lambda_{\mathrm{e}},\  x_1\ge 0}
\sum_{\sigma=\uparrow, \downarrow}
\sum_{j=1}^{\nu} \sum_{\vepsilon=\pm}'
(-t)  \Big(
e^{+i \alpha(\phi_{r(x)}-\phi_{r(x+\vepsilon \delta_j)})}
a_{r(x)\sigma}^* a_{r(x+\vepsilon \delta_j)\sigma}^*
+
\mathrm{h. c. }
\Big) \vartheta^{-1}\no
&=\vartheta 
\sum_{X\in \Lambda_{\mathrm{o}},\  X_1\le -1}
\sum_{\sigma=\uparrow, \downarrow}\sum_{j=1}^{\nu} \sum_{\vepsilon=\pm}'
(-t)  \Big(
e^{+i \alpha(\phi_X-\phi_{X+\vepsilon \delta_j})}
a_{X\sigma}^* a_{X +\vepsilon \delta_j\sigma}^*
+
\mathrm{h. c. }
\Big) \vartheta^{-1}\no
&=\vartheta 
\sum_{X\in \Lambda_{\mathrm{o}},\  X_1\le -1}
\sum_{\sigma=\uparrow, \downarrow}\sum_{j=1}^{\nu} \sum_{\vepsilon=\pm}'
(+t)  \Big(
e^{-i \alpha(\phi_{X+\vepsilon \delta_j}-\phi_X)}
 a_{X +\vepsilon \delta_j\sigma}^* a_{X\sigma}^*
+
\mathrm{h. c. }
\Big) \vartheta^{-1}\no
&=\vartheta T_L'' \vartheta^{-1}.
\end{align} 
Here, we use the fact that $r$ maps even sites  to odd
sites; namely,  if $x\in \Lambda_{\mathrm{e}}$, then $r(x) \in
\Lambda_{\mathrm{o}}$. Additionally, recall that  $\vartheta $ is antilinear.
 $\Box$
\medskip\\

The following lemmas  then immediately follow from (\ref{theta}).

\begin{lemm}\label{A2}
For all $\mb{h}_R\in \BbbR^{\Lambda_R}$, we define
 $r(\mb{h}_R)=\{h_{r^{-1}(x)}\}_{x\in \Lambda_R}\in \BbbR^{\Lambda_L} $. We have the
 following:
\begin{itemize}
\item[{\rm (i)}]
$
P''_R(\mb{h}_R)=\vartheta P''_L(r(\mb{h}_R))\vartheta^{-1}.
$ 
\item[{\rm (ii)}]
\begin{align}
P''_{LR}(\mb{h}) =&
-V \sum_{x\in \Lambda_{\mathrm{e}},\ x_1=0} (q_{x-\delta_1}-h_{x-\delta_1})
\otimes \vartheta (q_{x-\delta_1}-h_x)\vartheta^{-1}\no
&-V \sum_{x\in \Lambda_{\mathrm{e}},\ x_1=L-1} (q_{x+\delta_1}-h_{x+\delta_1})\otimes \vartheta(q_{x+\delta_1}-h_x)\vartheta^{-1}.
\end{align} 
\end{itemize} 
\end{lemm} 

\begin{lemm}\label{KTheta}
We have 
$
K_R=\vartheta K_L \vartheta^{-1}
$.
\end{lemm}

\begin{Prop}\label{RPZ}
For all $\mb{h}=(\mb{h}_L, \mb{h}_R)\in \BbbR^{\Lambda}$, 
let $Z_{\beta}(\mb{h})=Z_{\beta}(\mb{h}_L, \mb{h}_R)=\Tr[e^{-\beta H''_{\Lambda}(\mb{h})}]$.
Then we have 
\begin{align}
Z_{\beta}\big(\mb{h}_L, \mb{h}_R\big)^2\le Z_{\beta}\big(\mb{h}_L, r^{-1}(\mb{h}_L)\big)
Z_{\beta}\big(r(\mb{h}_R), \mb{h}_R\big), \label{LR}
\end{align} 
where $r^{-1}(\mathbf{h}_L):=\{h_{r(x)}\}_{x\in \Lambda_L}\in
 \BbbR^{\Lambda_R}$ for each $\mathbf{h}_L\in \BbbR^{\Lambda_L}$.
\end{Prop} 
{\it Proof.} Set 
\begin{align}
A&=T''_L+P''_L(\mb{h}_L)+K_L,\ \ \ B=T''_L+P''_L(r(\mb{h}_R))+K_L,\\
C_{x, \pm}^{(1)}&=q_{x\pm \delta_1}-h_{x\pm \delta_1},\ \ \ D_{x, \pm}^{(1)}=q_{x\pm
 \delta_1}-h_x,\\
C_{x, \pm}^{(2)}&=D_{x, \pm}^{(2)}=e^{i\alpha \phi_{x\pm \delta_1}}
 a_{x\pm \delta_1 \sigma}^*
\end{align} 
By Lemmas \ref{A1}--\ref{KTheta}, we see that $H''_{\Lambda}(\mb{h})$
has the form 
\begin{align}
A\otimes \one +\one \otimes \vartheta B \vartheta^{-1}
-  \sum_{x, \vepsilon, \sigma, \mu} \lambda_{x, \vepsilon, \mu}( C_{x,
\vepsilon}^{(\mu)}\otimes
\vartheta  D_{x, \vepsilon}^{(\mu)} \vartheta^{-1}
+ C_{x, \vepsilon}^{(\mu)*} \otimes  \vartheta D_{x, \vepsilon}^{(\mu)*}
\vartheta^{-1}
)
\end{align} 
 with $
 \lambda_{x, \vepsilon, \mu}\ge 0
$. Thus, we can apply Theorem \ref{DLS}. $\Box$
\medskip\\

\begin{coro}\label{Gauss}
For all $\mb{h}\in \BbbR^{\Lambda}$, we have 
$
Z_{\beta}(\mb{h}) \le Z_{\beta}(\mb{0}).
$
\end{coro} 
{\it Proof.}
Here, we give a sketch of the proof only. 
First, let us clarify the intuitive  meaning of  inequality
(\ref{LR}), namely, that 
the configurations
$(\mb{h}_L,r^{-1}(\mb{h}_L))$ and $(r(\mb{h}_R), \mb{h}_R)$ are more
``aligned'' than the original configuration  $\mb{h}=(\mb{h}_L,
\mb{h}_R)$
because $(r(\mb{h}_R), \mb{h}_R)$ and $(\mb{h}_L, r^{-1}(\mb{h}_L))$ are
invariant under   $r$-reflection.

For simplicity, assume that $
 Z_{\beta}\big(\mb{h}_L, r^{-1}(\mb{h}_L)\big)>
Z_{\beta}\big(r(\mb{h}_R), \mb{h}_R\big).
$
Proposition \ref{RPZ} concerns the reflection map $r$ with respect to a 
plane  $x_1=-1/2$ .
However, many reflection maps with respect to other
 planes exist as well. Thus, we can  apply  similar arguments
associated with another reflection map to $Z_{\beta}(\mb{h}_L,
r^{-1}(\mb{h}_L))$ and obtain an inequality similar to (\ref{LR}):
$Z_{\beta}(\mb{h}_L, r^{-1}(\mb{h}_L))^2 \le Z_{\beta}(\mb{h}_1)Z_{\beta}(\mb{h}_2)$.
 The key point is  that the  resulting  configurations $\mb{h}_1$ and $\mb{h}_2$  are  more aligned than $(\mb{h}_L,
r^{-1}(\mb{h}_L))$.
Repeating these procedures, we finally arrive at the most aligned
 configuration $\mb{h}_0=\mathrm{const}$. Since
$Z_{\beta}(\mb{h}_0)=Z_{\beta}(\mb{0})$, we obtain the  desired result (see
 \cite{DLS, FILS} for details). $\Box$

\subsubsection{Infrared bound}
Let $\Delta$ be the discrete Laplacian on $\Lambda$ given by 
\begin{align}
(\Delta \mb{h})_x=\sum_{j=1}^{\nu}(h_{x+\delta_j}+h_{x-\delta_j})- 2\nu h_x
\end{align} 
for all $\mb{h}=\{h_x\}_{x\in \Lambda}\in \BbbC^{\Lambda}$.
The following quantities will play essential roles:
\begin{align}
g &=\Big\la \la \mb{q}|(-\Delta) \mb{h}\ra^* 
\la \mb{q}| (-\Delta) \mb{h}\ra\Big\ra_{\beta, \Lambda}'',\\
b&= \Big(
\la \mb{q}| (-\Delta) \mb{h}\ra,  
\la \mb{q}| (-\Delta) \mb{h}\ra
\Big)_{\beta, \Lambda}'',\\
c&= \beta
\Big\la
\Big[
\la \mb{q}| (-\Delta) \mb{h}\ra, \Big[
H''_{\Lambda},
\la \mb{q}| (-\Delta) \mb{h}\ra^*
\Big]
\Big]
\Big\ra_{\beta, \Lambda}''.
\end{align} 
Here, we used the following notations:
\begin{itemize}
\item $\la\cdot \ra''_{\beta, \Lambda}$ is the thermal expectation
      associated with $H_{\Lambda}''$.
\item  $
(A, B)_{\beta, \Lambda}''
$
 is the Duhamel two-point function associated with
       $H_{\Lambda}''$. Namely, 
\begin{align}
(A, B)_{\beta, \Lambda}''&=(Z_{\beta}'')^{-1}\int_0^1dx \Tr\big[e^{-x\beta
 H_{\Lambda}''}
A^* e^{-(1-x) \beta H_{\Lambda}''} B
\big],\\ Z_{\beta}''&=\Tr\big[e^{-\beta H_{\Lambda}''}\big].
\end{align} 
\item 
$\la \mb{q}|(-\Delta) \mb{h}\ra=\sum_{x\in \Lambda} q_x (-\Delta \mb{h})_x$.
\end{itemize} 
We begin with the following lemma:
\begin{lemm}\label{LemmaB}
For all  $\mb{h}\in \BbbC^{ \Lambda}$, we have 
\begin{align}
b\le b_0,\label{Infra}
\end{align}
where $ \displaystyle b_0=\frac{(\beta V)^{-1}}{2} \la \mb{h}|
 (-\Delta)\mb{h}\ra$.  Here, $\la \cdot| \cdot \ra$ is the standard
 inner product on $\BbbC^{\Lambda}$.
\end{lemm} 
{\it Proof.}  By Corollary \ref{Gauss}, we have 
$
d^2Z_{\beta}(\lambda \mb{h})\Big/d\lambda^2\Big|_{\lambda=0}\le 0
$. Thus, we obtain (\ref{Infra}) for all $\mb{h}\in \BbbR^{\Lambda}$.
Using the well-known fact
$
(A^*, A)_{\beta, \Lambda}''=(A_R, A_R)_{\beta, \Lambda}''+(A_I, A_I)_{\beta,
\Lambda}''
$
 with $A_R=(A+A^*)/2$ and $A_I=(A-A^*)/2i$, we can extend (\ref{Infra})
 to complex $\mb{h}$. $\Box$

\begin{lemm}\label{LemmaC}
Let $\tau$ be a unitary operator on $\BbbC^{\Lambda}$ given by $(\tau
 \mb{h})_x=\{(-1)^{\|x\|} h_x\}_{x\in \Lambda}$.
For all $\mb{h}\in \BbbC^{\Lambda}$, we have 
\begin{align}
c\le
c_0
\end{align} 
where $c_0=
 4\beta t\Big \la (-\Delta) \mb{h}\Big| \tau (-\Delta)\tau^{-1} (-\Delta) \mb{h}\Big\ra.
$
\end{lemm} 
{\it Proof.} Let $M=\tau(-\Delta)$. By direct computation, we have
\begin{align*}
&\Big[
\la \mb{q}| (-\Delta) \mb{h}\ra, \Big[
H''_{\Lambda},
\la \mb{q}| (-\Delta) \mb{h}\ra^*
\Big]
\Big]\no
&=\sum_{\la x; y\ra} \sum_{\sigma=\uparrow, \downarrow}(-t) 
\Big|
(M\mb{h})_x-(M\mb{h})_y
\Big|^2
\big(
e^{-i\alpha (\phi_x-\phi_y)}c_{x\sigma}^*c_{y\sigma}^*+{\mathrm{h.c.}}
\big).
\end{align*} 
Hence, we have
$
c\le 4t\beta \la (M\mb{h})| (-\Delta)(M\mb{h})\ra=c_0.
$ 
This completes the proof. $\Box$

\begin{Prop}\label{PropG}
For all $\mb{h}\in \BbbC^{\Lambda}$, we have 
\begin{align}
g\le 
(2\pi)^{\nu/2}
\gamma_1
\la \mb{h}|(-\Delta)\mb{h}\ra+(2\pi)^{-\nu/2} \gamma_2 
\big\la
 (-\Delta)\mb{h}|\tau^{-1} (-\Delta) \tau (-\Delta)\mb{h}
\big\ra.\label{GINQ}
\end{align} 
\end{Prop} 
{\it Proof.} Applying the Falk--Bruch inequality \cite{FB, Simon2}, we have 
\begin{align}
g\le \frac{1}{2}\sqrt{bc} \coth \sqrt{\frac{c}{4b}}.
\end{align} 
Since $\displaystyle 
\coth x\le 1+\frac{1}{x}$, we have 
\begin{align}
g\le \frac{1}{2}\sqrt{bc}+b.
\end{align} 
By Lemmas \ref{LemmaB} and \ref{LemmaC},  it holds that  
\begin{align}
\sqrt{bc}\le \Big(\frac{t}{V}\Big)^{1/2} \la \mb{h}|(-\Delta) \mb{h}\ra
+\frac{1}{2}\Big(
\frac{t}{V}
\Big)^{1/2}\big\la
 (-\Delta)\mb{h}|\tau^{-1} (-\Delta) \tau (-\Delta)\mb{h}
\big\ra.
\end{align} 
Thus, we obtain the desired result. $\Box$
\medskip\\

\begin{Thm}
Let $\displaystyle
(2\pi)^{-\nu /2}c_{\beta}=\liminf_{\|x\|\to \infty} \la q_x q_o\ra_{\beta}''.
$
We have 
\begin{align}
\la q_o^2\ra_{\beta}'' \le (2\pi)^{-\nu/2}c_{\beta}+\gamma_1\int_{\mathbb{T}^{\nu}}
 \frac{dp}{E(p)}
+\gamma_2.\label{qBound}
\end{align} 
\end{Thm} 
{\it Proof.} Let $G_{\beta, \Lambda}(x-y)=\la q_x q_y\ra_{\beta, \Lambda}''$.
Let $E(p)=\sum_{j=1}^{\nu}(1-\cos p_j)$ and let $F(p)=\sum_{j=1}^{\nu}(1+\cos p_j)$.
By the Fourier transformation,  we see that 
\begin{align}
g&= \frac{(2\pi)^{\nu}}{|\Lambda|}
\sum_{p\in B_L}E(p)^2 \Big\{
(2\pi)^{\nu/2} \hat{G}_{\beta, \Lambda}(p)
\Big\} |\hat{h}(p)|^2,\\
\la \mb{h}|(-\Delta) \mb{h}\ra&=
\frac{(2\pi)^{\nu}}{|\Lambda|} \sum_{p\in B_L}E(p) |\hat{h}(p)|^2,\\
\big\la
 (-\Delta)\mb{h}|\tau^{-1} (-\Delta) \tau (-\Delta)\mb{h}
\big\ra&=
\frac{(2\pi)^{\nu}}{|\Lambda|} \sum_{p\in B_L}E(p)^2F(p) |\hat{h}(p)|^2,
\end{align} 
where
$\hat{h}(p)=(2\pi)^{-\nu/2} \sum_{x\in \Lambda} e^{-i x\cdot p} h_x$.

By inserting these formulas into (\ref{GINQ}) and taking $L\to \infty$,
we obtain 
\begin{align}
&\int_{\mathbb{T}^{\nu}} dp E(p)^2 (2\pi)^{\nu/2} \hat{G}_{\beta}(p)
 |\hat{h}(p)|^2\no
&\le (2\pi)^{\nu/2}\gamma_1 \int_{\mathbb{T}^{\nu}}dpE(p) |\hat{h}(p)|^2
+(2\pi)^{-\nu/2}\gamma_2 \int_{\mathbb{T}^{\nu}}dp F(p) E(p)^2|\hat{h}(p)|^2.
\end{align} 
From this, we know that $\hat{G}_{\beta}$ has the following form
\begin{align}
\hat{G}_{\beta}(p)=a_{\beta} \delta(p)+I_{\beta}(p),
\end{align} 
where $I_{\beta}(p)$ satisfies
\begin{align}
I_{\beta}(p)\le \gamma_1\frac{1}{E(p)}+(2\pi)^{-\nu} \gamma_2 F(p).\label{IInq}
\end{align}  
On the other hand, we have
\begin{align}
\la q_o^2\ra_{\beta}''=(2\pi)^{-\nu/2}\int_{\mathbb{T}^{\nu}} dp\, \hat{G}_{\beta}(p)
=(2\pi)^{-\nu/2}a_{\beta}+(2\pi)^{-\nu /2}\int_{\mathbb{T}^{\nu}}dp I_{\beta}(p). \label{ImpEq}
\end{align} 
Combining this with (\ref{IInq}), we obtain 
\begin{align}
\la q_o^2\ra_{\beta}'' \le (2\pi)^{-\nu/2}a_{\beta}+\gamma_1\int_{\mathbb{T}^{\nu}}
 \frac{dp}{E(p)}
+\gamma_2.
\end{align}

Finally, we show that $a_{\beta}=c_{\beta}$.
To this end, we observe that  
\begin{align}
\la q_x q_o\ra_{\beta}''= (2\pi)^{-\nu /2}
\int_{\mathbb{T}^{\nu}} dp e^{-i x \cdot p} \hat{G}_{\beta}(p)
=(2\pi)^{-\nu /2} a_{\beta}+(2\pi)^{-\nu /2}
 \int_{\mathbb{T}^{\nu}} dp\,  e^{ip\cdot x} I_{\beta}(p). 
\end{align} 
By the Riemann--Lebesgue lemma, we know that 
$\displaystyle 
\lim_{\|x\|\to \infty} \int_{\mathbb{T}^{\nu }} dp\,  e^{ix \cdot p } I_{\beta}(p)=0
$. Thus, we have 
$a_{\beta}=c_{\beta}$. $\Box$

\subsection{Lower bound for $\la q_o^2\ra_{\beta}''$}

\begin{lemm}\label{InqM}
Assume that $\nu V-U_{\mathrm{eff}}>0$.
Let $M_{\Lambda}=H''_{\Lambda}-K$.
 We have 
\begin{align}
\Big\la -\frac{M_{\Lambda}}{|\Lambda|}\Big\ra_{\beta, \Lambda}''
\le 8\nu t +(\nu V-U_{\mathrm{eff}})\la q_o^2\ra_{\beta, \Lambda}''.
\end{align} 
\end{lemm} 
{\it Proof.}
First, remark that 
\begin{align}
-M_{\Lambda}=-T''+(\nu V-U_{\mathrm{eff}}) \sum_{x\in \Lambda}q_x^2-\frac{1}{2}
\sum_{\la x; y\ra}V(q_x-q_y)^2.
\end{align} 
Since
\begin{align}
\Bigg\la \sum_{\la x;y\ra}(q_x-q_y)^2\Bigg\ra_{\beta,\Lambda}''&\ge 0,\\
 \la q_x^2\ra_{\beta, \Lambda}''&=\la q_o^2\ra_{\beta, \Lambda}'', \ \forall
 x\in \Lambda,\\
\big| \la T'' \ra_{\beta, \Lambda}'' \big|&\le 2t \sum_{\la x; y\ra}
 \sum_{\sigma} 1=8\nu t |\Lambda|,
\end{align} 
we obtain the assertion in the lemma. $\Box$

\begin{lemm}\label{InqTr}
We have 
\begin{align}
\ln \Tr[e^{-\beta H''_{\Lambda}}] \ge \beta (\nu V-U_{\mathrm{eff}})|\Lambda|. 
\end{align} 
\end{lemm} 
{\it Proof.}
Let 
\begin{align}
\Psi=\Bigg[\prod_{x\in \Lambda}
 c_{x\uparrow}^*c_{x\downarrow}^*\tilde{\Omega}_{\mathrm{f}}\Bigg]\otimes 
\tilde{\Omega}_{\mathrm{b}},
\end{align} 
where $\tilde{\Omega}_{\mathrm{f}}$
(resp. $\tilde{\Omega}_{\mathrm{b}}$) is the Fock vacuum in $\Fock$
(resp. $\mathfrak{P}$).
Then,  we otain 
\begin{align}
\la \Psi|H''_{\Lambda}\Psi\ra=-(\nu V-U_{\mathrm{eff}})|\Lambda|.
\end{align} 
By  the Peierls--Bogoliubov inequality \cite{Simon2}, we have  
\begin{align}
\Tr[e^{-\beta H''_{\Lambda}}] \ge  e^{-\beta \la
 \Psi|H''_{\Lambda}\Psi\ra}
=e^{\beta (\nu V-U_{\mathrm{eff}})|\Lambda|}.
\end{align} 
This completes the proof. $\Box$

\begin{Prop}
We have 
\begin{align}
\la q_o^2\ra_{\beta, \Lambda}''\ge 1
-8\nu t ( \nu V-U_{\mathrm{eff}})^{-1} -(\nu V-U_{\mathrm{eff}})^{-1} \ln
 4(1-e^{-\beta  \omega})^{-1}. \label{LowerB}
\end{align} 
\end{Prop} 
{\it Proof.} By Lemma \ref{UsefulLemm}, we have
\begin{align}
\ln \Tr[e^{-\beta H''_{\Lambda}}]\le \la (-\beta
 M_{\Lambda})\ra_{\beta, \Lambda}''+\ln \Tr[e^{-\beta K}].
\end{align} 
Note that 
$
\Tr[e^{-\beta K}]=4^{|\Lambda|} (1-e^{-\beta \omega})^{-|\Lambda|}
$. Combining this with Lemmas \ref{InqM} and \ref{InqTr}, we obtain the desired
result. $\Box$

\subsection{Completion of proof of Theorem \ref{Main}}

By (\ref{qBound}), (\ref{LowerB}),
and  $\la q_x q_o\ra_{\beta}''=(-1)^{\|x\|}\la q_x q_o\ra_{\beta} $
, we obtain the assertion in the theorem. $\Box$

\appendix

\section{Proof of (\ref{Half})}\label{PfHalf}
\setcounter{equation}{0}
We will show that  $\la q_x\ra_{\beta, \Lambda}=0$.
The {\it hole-particle  transformation} is a unitary operator $u$ such that 
\begin{align}
u c_{x\uparrow} u^{-1} =
c_{x\uparrow},\ \ \ u c_{x\downarrow} u^{-1} =(-1)^{\|x\|}c_{x\downarrow}^*.
\end{align}
We set $\mathbb{H}=u H_{\Lambda} u^{-1}$. Since 
$uq_xu^{-1}=s_x:=n_{x\uparrow}-n_{x\downarrow}$, we obtain $\mathbb{H}=\mathbb{H}_0+\mathbb{W}$, where
\begin{align}
\mathbb{H}_0&=T+K,\\
\mathbb{W} &= U\sum_{x\in \Lambda} s_x^2+V\sum_{\la x, y\ra}s_xs_y+g\sum_{x\in \Lambda} s_x(b_x+b_x^*).
\end{align}
Here, $T$ and $K$ are defined by (\ref{EKinetic}) and (\ref{BKinetic}), respectively.
Thus, we have
\begin{align}
\la q_x\ra=\la\!\la s_x\ra\!\ra,
\end{align}
where $\la\!\la \cdot \ra\!\ra$ is the thermal expectation associated with $\mathbb{H}$.

Let $D$ be a   unitary operator such that
\begin{align}
Dc_{x\uparrow} D^{-1} =c_{x\downarrow},\ \ D c_{x\downarrow} D^{-1}=c_{x\uparrow}, \ \ Db_x D^{-1}=-b_x.
 \end{align} 
Since $D\mathbb{H} D^{-1}=\mathbb{H}$ and $Ds_x D^{-1} =-s_x$,  we have
 $\la q_x\ra=\la \!\la s_x\ra\!\ra=0$.
This concludes the proof of (\ref{Half}). $\Box$

\section{The Dyson--Lieb--Simon inequality} \label{AppA}
\setcounter{equation}{0}

Let $\mathfrak{X}_L$ and $\mathfrak{X}_R$ be   complex Hilbert spaces and let $\vartheta$ be an
antiunitary  transformation from  $\mathfrak{X}_L$ onto $\mathfrak{X}_R$. 
Let $A, B, C_j, D_j, j=1,\dots, n$ be  linear operators in $\mathfrak{X}_L$.
Suppose that $A$ and $B$ are self-adjoint and  bounded from below and that $C_j$ and $D_j$ are bounded.
We will  study the following Hamiltonian:
\begin{align}
H(A, B, \mathbf{C}, \mathbf{D})&=H_0-V,\\
H_0&=A\otimes \one+\one \otimes \vartheta B \vartheta^{-1},\\
V&= \sum_{j=1}^n \lambda_j(C_j\otimes \vartheta D_j\vartheta^{-1}+C_j^*\otimes
 \vartheta D_j^* \vartheta^{-1}).
\end{align} 
$H(A, B, \mathbf{C}, \mathbf{D})$ is a self-adjoint operator bounded
from below   acting  in $\mathfrak{X}_L\otimes \mathfrak{X}_R$. 
\begin{Thm}\label{DLS}
Assume that $e^{-\beta A}$ and $e^{-\beta B}$ are trace class operators
 for all $\beta >0$ and that $\lambda_j\ge 0$ for all $j\in \{1,
 \dots, n\}$.
Let
$
Z_{\beta}(A, B, \mathbf{C}, \mathbf{D})
=\mathrm{Tr}
\big[
e^{-\beta H(A, B, \mathbf{C}, \mathbf{D})}
\big],\ \beta >0
$. We  then have 
\begin{align}
Z_{\beta}(A, B, \mathbf{C}, \mathbf{D})^2 \le Z_{\beta}(A, A,
 \mathbf{C}, \mathbf{C})Z_{\beta}(B, B, \mathbf{D}, \mathbf{D}).
\end{align} 
\end{Thm} 
\begin{rem}{\rm 
\begin{itemize}
\item[{\rm (i)}] In  \cite{DLS},  all matrix elements of $A, B, C_j, D_j$ are assumed to be
 {\it real}. However, as noted in \cite{LiebNach, Miyao1}, this assumption is unnecessary.
This point is essential for  the present  paper. 
\item[{\rm (ii)}]
Suppose that  $\dim \mathfrak{X}_L<\infty$. Therefore, we set $\mathfrak{X}_L=\mathfrak{X}_R=\BbbC^n$.
Let $\vartheta$ be the standard conjugation: $\vartheta \psi=\{\overline{\psi_j}\}_{j=1}^n$ for each $ \psi\in \mathfrak{X}_L$.
 Hence,   $\vartheta B\vartheta^{-1}$
 represents the   complex conjugation of the matrix elements of $B$.
Now assume  the following: (a) $C_j=D_j$ for all $j$; (b) $C_j$ is self-adjoint for all $j$ ($C_j^*=C_j$);
(c) $C_j$ is real for all $j$ ($\vartheta C_j \vartheta^{-1}=C_j$).  In this case, we obtain a finite temperature version of     \cite[Lemma 14]{LiebNach}. 
$\diamondsuit$
\end{itemize}
}
\end{rem} 
{\it Proof.} 
While this  theorem is proven  in \cite{Miyao1},  we present the  
proof here   for reader' convenience. It suffices to show the assertion when $\dim \mathfrak{X}_L<\infty$.

The following property is fundamental:
\begin{align}
\Tr_{\mathfrak{X}_L\otimes \mathfrak{X}_R}\Big[
A\otimes \vartheta B\vartheta^{-1}
\Big]=
\Tr_{\mathfrak{X}_L}[A] \big(
\Tr_{\mathfrak{X}_L}[B]
\big)^*. \label{BASIC}
\end{align} 
Especially, we have $
\Tr_{\mathfrak{X}_L\otimes \mathfrak{X}_R}\Big[
A\otimes \vartheta A\vartheta^{-1}
\Big]=
\big|\Tr_{\mathfrak{X}_L}[A]\big|^2\ge 0
$. The reason for this  is as follows.
Since 
$
\Tr_{\mathfrak{X}_L\otimes \mathfrak{X}_R}\Big[
A\otimes \vartheta B\vartheta^{-1}
\Big]=
\Tr_{\mathfrak{X}_L}[A] 
\Tr_{\mathfrak{X}_R}[\vartheta B \vartheta^{-1}]
$, it suffices to show that $
\Tr_{\mathfrak{X}_R}[\vartheta B \vartheta^{-1}]
=(\Tr_{\mathfrak{X}_L}[ B ])^*
$.
Let $\{e_n\}_{n=1}^{\infty}$ be a complete orthonormal system (CONS) of
$\mathfrak{X}_R$. Remarking that $\{\vartheta^{-1} e_n\}_{n=1}^{\infty}$
is a CONS of $\mathfrak{X}_L$ and since $\la \vartheta^{-1}
\phi|\vartheta^{-1} \psi\ra=\la \psi|\phi\ra$, we see that 
\begin{align}
\Tr_{\mathfrak{X}_R}[\vartheta B \vartheta^{-1}]&=\sum_{n=1}^N \la
 e_n|\vartheta B\vartheta^{-1}e_n\ra
=\sum_{n=1}^N  
 \la \vartheta^{-1}\vartheta B\vartheta^{-1}e_n|\vartheta^{-1}e_n\ra\no
&=\sum_{n=1}^N \la B\vartheta^{-1} e_n|\vartheta^{-1} e_n\ra
=(\Tr_{\mathfrak{X}_L}[B])^*. \label{CTr}
\end{align}

As a first step, we will prove the assertion by  assuming 
 that $C_j$ and $D_j$ are self-adjoint. 
For simplicity,  assume
 that $\lambda_j=1/2$. 
By the Duhamel formula, 
\begin{align}
\ex^{-\beta H(A, B; \mathbf{C}, \mathbf{D})}&=\sum_{N\ge 0}\mathscr{D}_{N, \beta}(A, B; \mathbf{C}, \mathbf{D}),\label{Duha}\\
\mathscr{D}_{N, \beta}(A, B; \mathbf{C}, \mathbf{D})&=\int_{S_N(\beta)}\
 e^{-t_1 H_0}V e^{-t_2 H_0}\cdots  e^{-t_N
 H_0}V e^{-(\beta-\sum_{j=1}^N t_j)
 H_0},
\label{DUHA2}
\end{align}
where $\int_{S_N(\beta)}=\int_0^{\beta}\dm t_1\int_0^{\beta-t_1}\dm
t_2\cdots \int_0^{\beta-\sum_{j=1}^{N-1}t_j} \dm t_N
$.  
Observe that 
\begin{align}
&
\mathscr{D}_{N, \beta}(A, B; \mathbf{C}, \mathbf{D})
\no
=&\sum_{k_1, \dots, k_N\ge 1}
\int_{S_N(\beta)}\ 
\Big[\mathscr{L}_{A; \mathbf{C}}\big(\mathbf{k}_{(N)}; \mathbf{t}_{(N)}\big)
\Big]
\otimes 
 \vartheta  \Big[\mathscr{L}_{B; \mathbf{D}}\big(\mathbf{k}_{(N)}; \mathbf{t}_{(N)}\big)
\Big] \vartheta^{-1}, \label{BasicStr}
\end{align} 
where $\mathbf{k}_{(N)}=(k_1,\dots, k_N)\in \BbbN^N,$ $\mathbf{t}_{(N)}=(t_1,
\dots, t_N)\in \BbbR^N_+$ and 
\begin{align}
\mathscr{L}_{X; \mathbf{Y}}\big(\mathbf{k}_{(N)};
 \mathbf{t}_{(N)}\big)=e^{-t_1 X}Y_{k_1}e^{-t_2 X}\cdots e^{-t_N
 X}Y_{k_N} e^{-(\beta-\sum_{j=1}^N t_j)X}
\end{align} 
 with $\mathbf{Y}=\{Y_j\}_j$. 
By this fact and (\ref{CTr}), we observe that 
\begin{align}
&\Tr_{\mathfrak{X}_L\otimes \mathfrak{X}_R}\Big[
\mathscr{D}_{N, \beta}(A, B; \mathbf{C}, \mathbf{D})
\Big]\no
=&\sum_{k_1, \dots, k_N\ge 1}\int_{S_N(\beta)}
\Big\{\Tr_{\mathfrak{X}_L}\Big[\mathscr{L}_{A; \mathbf{C}}\big(\mathbf{k}_{(N)};
 \mathbf{t}_{(N)}
\big)
\Big]
\Big\}
\times \Big\{\Tr_{\mathfrak{X}_L}\Big[\mathscr{L}_{B; \mathbf{D}}\big(\mathbf{k}_{(N)}; \mathbf{t}_{(N)}\big)
\Big]
\Big\}^*.
\end{align} 
Let us introduce an  inner product by 
\begin{align}
\la F| G\ra_{N, \beta}=\sum_{k_1, \dots, k_N\ge
 1}\int_{S_N(\beta)}\,
F\big(\mathbf{k}_{(N)}; \mathbf{t}_{(N)}\big)G\big( \mathbf{k}_{(N)};
 \mathbf{t}_{(N)}\big)^*.\label{InnerProduct}
\end{align} 
In terms of this inner product, we have 
\begin{align}
&\Tr_{\mathfrak{X}_L\otimes \mathfrak{X}_R}\Big[
 \mathscr{D}_{N, \beta}(A, B; \mathbf{C}, \mathbf{D})
\Big]=\Big\la F_{A;\mathbf{C}}^{(N)}\Big| F_{B;\mathbf{D}}^{(N)}\Big\ra_{N, \beta},
\end{align} 
where 
\begin{align}
F_{X;\mathbf{Y}}^{(N)}\big(\mathbf{k}_{(N)};
 \mathbf{t}_{(N)}\big)=\Tr_{\mathfrak{X}_L}\Big[ \mathscr{L}_{X; \mathbf{Y}}
\big(\mathbf{k}_{(N)}; \mathbf{t}_{(N)}\big)\Big].
\end{align} 
By the Schwartz inequality, we have 
\begin{align}
\bigg|
\Tr_{\mathfrak{X}_L\otimes \mathfrak{X}_R}\Big[
e^{-\beta H(A, B, \mb{C}, \mb{D})}
\Big]
\bigg|^2
&= \Bigg|
\sum_{N\ge 0} \Big\la F_{A;\mathbf{C}}^{(N)}\Big| F_{B;\mathbf{D}}^{(N)}
\Big\ra_{N, \beta}\Bigg|^2 \no
&\le \Bigg(
\sum_{N\ge 0}\big \|F_{A;\mathbf{C}}^{(N)}\big\|^2_{N, \beta}
\Bigg)
 \Bigg(
\sum_{N\ge 0}\big\|F_{B; \mathbf{D}}^{(N)}\big\|^2_{N, \beta}
\Bigg), \label{Inq1}
\end{align}
where $\|W \|^2_{N, \beta}:=\la W| W \ra_{N, \beta}$. 
Finally,  we remark that 
\begin{align}
\sum_{N \ge 0}\big\|F_{A; \mathbf{C}}^{(N)}\big\|^2_{N, \beta}&=\sum_{N\ge 0}
\sum_{k_1, \dots, k_N\ge 1}\int_{S_N(\beta)} 
\bigg|\Tr_{\mathfrak{X}}\Big[ \mathscr{L}_{A; \mathbf{C}}\big(\mathbf{k}_{(N)};
 \mathbf{t}_{(N)}\big)
\Big]
\bigg|^2\no
&=\sum_{N\ge 0}\Tr_{\mathfrak{X}_L\otimes \mathfrak{X}_R}
\Big[
\mathscr{D}_{N, \beta}(A, A; \mathbf{C}, \mathbf{C})
\Big]
\no
&=\Tr_{\mathfrak{X}_L\otimes \mathfrak{X}_R}
\Big[
e^{-\beta H(A, A; \mathbf{C}, \mathbf{C})}
\Big].\label{Inq2}
\end{align} 
Combining (\ref{Inq1}) and (\ref{Inq2}), we obtain the assertion
for the case where  $C_j$ and $D_j$ are self-adjoint. 

We note that for general $C_j$ and $D_j$,  these operators can be written as 
\begin{align}
C_j=\Re C_j+i \Im C_j,\ \ D_j=\Re D_j+i \Im D_j,
\end{align} 
where $\Re C_j, \Re D_j, \Im C_j$ and $\Im D_j$ are self-adjoint.
Since
\begin{align}
C_j\otimes \vartheta D_j\vartheta^{-1}+C_j^*\otimes
 \vartheta D_j^* \vartheta^{-1}
= 2(\Re C_j\otimes \vartheta \Re D_j\vartheta^{-1}+\Im C_j\otimes \vartheta \Im D_j \vartheta^{-1}),
\end{align} 
we can reduce  the problem to  the case where $C_j$ and $D_j$ are self-adjoint.
$\Box$

\section{A useful lemma }\label{AppB}
\setcounter{equation}{0}
\begin{lemm} \label{UsefulLemm}
Let $B$ and  $C$ be self-adjoint operators. Suppose that 
$e^{-C}$ is a trace class operator and suppose that $B$ is bounded.
We have 
\begin{align}
\ln \Tr\big[e^{-(B+C)}\big] \le \la -B\ra+\ln \Tr\big[
e^{-C}
\big],
\end{align}  
where 
$
\la X \ra=\Tr\big[
X e^{-(B+C)}
\big]\Big/\Tr\big[
e^{-(B+C)}
\big]$.
\end{lemm} 
{\it Proof.} Let $X$ and $Y$ be self-adjoint. We know that 
$F(\lambda)=\ln \Tr[e^{\lambda X+(1-\lambda) Y}]$ is convex,  e.g.,
as in  \cite{Simon2}.
 Thus, we have $F(1) \ge F'(0)+F(0)$, which implies
\begin{align}
\ln \Tr[e^X]\ge \la X-Y\ra_Y+\ln \Tr[e^Y],
\end{align}  
where $\la L \ra_Y=\Tr[L\,  e^Y]\big/\Tr[e^Y]$.
Substituting  $X=-C$ and $Y=-B-C$, we obtain the desired result. $\Box$

\section{Proof of Proposition \ref{AntiUni}}\label{AntiU}
\setcounter{equation}{0}
Let 
\begin{align}
\mathscr{S}_n^{(0)}=
\big\{
(X_1,\dots, X_n)\, \big|\, X_j\in \Lambda\times \{\uparrow, \downarrow\},
\, j=1, \dots, n\  \mbox{and}\  X_i\neq X_j,\, \mbox{if}\  i\neq j 
\big\}.
\end{align}
Let $\mathfrak{S}_n$ be the permutation group on  set $\{1, \dots,
n\}$. 
Let $(X_1, \dots, X_n),\, (Y_1, \dots, Y_n)\in \mathscr{S}_n^{(0)}$.
If there exists a $\pi\in \mathfrak{S}_n$ such that 
$(X_{\pi(1)}, \dots, X_{\pi(n)})=(Y_1, \dots, Y_n)$, then we write $
(X_1, \dots, X_n) \sim (Y_1, \dots, Y_n)
$.
The binary relation ``$\sim$'' on $\mathscr{S}_n^{(0)}$ is an
equivalence relation. We denote  the quotient set
$\mathscr{S}_n^{(0)} \backslash \sim$ by $\mathscr{S}_n$ and 
for the simplicity of notation, we still denote
the equivalence class $[(X_1, \dots, X_n)]$  by $(X_1, \dots, X_n)$.

Set $\displaystyle
\mathscr{S}=\bigcup_{n=0}^{2|\Lambda|} \mathscr{S}_n
$, where $\mathscr{S}_{n=0}=\{\emptyset\}$. Let
\begin{align}
\mathscr{S}_L&=
\big\{
(X_1,\dots, X_n)\in \mathscr{S}\, \big|\, X_j\in \Lambda_L\times \{\uparrow, \downarrow\},
\, j=1, \dots, n,\ n\in \{0\} \cup\BbbN
\big\},\\
\mathscr{S}_R&=
\big\{
(X_1,\dots, X_n)\in \mathscr{S}\, \big|\, X_j\in \Lambda_L\times \{\uparrow, \downarrow\},
\, j=1, \dots, n,\ n\in \{0\}\cup\BbbN
\big\}.
\end{align}
Here, if $n=0$, then we understand that $(X_1, \dots, X_n)=\emptyset$.
For each $X=(x, \sigma)\in \Lambda\times \{\uparrow, \downarrow\}$, we set
$c_X:=c_{x\sigma}$ and $a_X:=a_{x\sigma}$.
For each $\mathbf{X}=(X_1, \dots, X_n)\in \mathscr{S}$, we define
\begin{align}
e(\mathbf{X})=c_{X_1}^*\dots c_{X_n}^* \Omega_{\mathrm{f}},\ \ \  
f(\mathbf{X})=a_{X_1}^*\cdots a_{X_n}^*\Omega_{\mathrm{f}}, \label{CONSs}
\end{align}
and  $e(\emptyset)=\Omega_{\mathrm{f}}$,
$f(\emptyset)=\Omega_{\mathrm{f}}$.
The definition (\ref{CONSs}) is independent of  the  choice of
the representative up to the sign factor,  and 
trivially, $\{e(\mathbf{X})\, |\, \mathbf{X}\in \mathscr{S}_R\}$ is a  CONS of $\Fock_R$.
We note that $\{a_X\, |X\in \Lambda\times \{\uparrow, \downarrow\}\}$ satisfies the CARs:
\begin{align}
\{a_X, a_Y^*\}=\delta_{XY},\ \ \ \{a_X, a_Y\}=0.
\end{align}
Moreover, it holds that $a_X\Omega_{\mathrm{f}}=0$ for all $X\in \Lambda\times \{\uparrow, \downarrow\}$.
Thus, $\{f(\mathbf{X})\, |\, \mathbf{X}\in \mathscr{S}_L\}$ is a CONS of $\Fock_L$.

For each $X=(x, \sigma)\in \Lambda_R\times \{\uparrow, \downarrow\}$,
we set $r(X):=(r(x), \sigma)\in \Lambda_L\times \{\uparrow,
\downarrow\}$, where $r$  in the right-hand side 
is defined by (\ref{ReflectionMap}).
For each $\mathbf{X}=(X_1, \dots, X_n)\in \mathscr{S}_R$, we further extend the map $r$ as follows:
\begin{align}
r(\mathbf{X}):=(r(X_1), \dots, r(X_n))\in \mathscr{S}_L.
 \end{align} 
Thus, $\{f(r(\mathbf{X}))\, |\, \mathbf{X}\in \mathscr{S}_R\}$ is a CONS
of $\Fock_L$.
For each $\Psi\in \Fock_L$, we have the following expression:
\begin{align}
\Psi=\sum_{\mathbf{X}\in \mathscr{S}_R}
\Psi(r(\mathbf{X})) f(r(\mathbf{X})),\ \ \ \ 
\Psi(r(\mathbf{X}))=\la f(r(\mathbf{X}))|\Psi\ra. \label{ExPhi}
\end{align}
Using the expression (\ref{ExPhi}), we define  
 an antilinear map $\xi$ from $\Fock_L$ onto $\Fock_R$  by 
\begin{align}
\xi\Psi=\sum_{\mathbf{X}\in \mathscr{S}_R} \overline{\Psi(r(\mathbf{X}))} e(\mathbf{X})\label{Xi1}
\end{align}
and $\xi \Omega_{\mathrm{f}}^L=\Omega_{\mathrm{f}}^R$. 
$\xi^{-1}$ is given by 
\begin{align}
\xi^{-1} \Phi=\sum_{\mathbf{X}\in \mathscr{S}_R}
 \overline{\Phi(\mathbf{X})} f(r(\mathbf{X})) \label{Xi2}
\end{align}
for each $\Phi=\sum_{X\in \mathscr{S}_R} \Phi(\mathbf{X})
e(\mathbf{X})\in \Fock_R$.
It  is not difficult to check that $
\la \xi\Psi_1|\xi\Psi_2\ra=\overline{\la \Psi_1|\Psi_2\ra}
$ for all $\Psi_1, \Psi_2\in \Fock_L$.
Hence, $\xi$ is an antiunitary transformation.

\begin{lemm}\label{Xi}
For all $X\in \Lambda_R\times \{\uparrow, \downarrow\}$, it holds that 
$\xi a_{r(X)}\xi^{-1}=c_X$.
\end{lemm}
{\it Proof.}
For each $\Phi=\sum_{\mathbf{Y}\in \mathscr{S}_R}
\Phi(\mathbf{Y})e(\mathbf{Y})\in \Fock_R$, we have, by (\ref{Xi1}) and
(\ref{Xi2}),
\begin{align}
\xi a_{r(X)}^* \xi^{-1} \Phi
&=\xi a_{r(X)}^* \sum_{\mathbf{Y}\in \mathscr{S}_R}
\overline{\Phi(\mathbf{Y})}f(r(\mathbf{Y}))
=\xi \sum_{\mathbf{Y}\in \mathscr{S}_R,\  X\notin \mathbf{Y}}
\overline{\Phi(\mathbf{Y})}f(r(X, \mathbf{Y}))\no
&=\sum_{\mathbf{Y}\in \mathscr{S}_R,\  X\notin \mathbf{Y}}
\Phi(\mathbf{Y})e(X, \mathbf{Y})
=c_X^* \Phi.
\end{align} 
Hence, $\xi a_{r(X)}^* \xi^{-1}=c_X^*$. $\Box$
\medskip\\

Recall that $\mathfrak{H}_L=\Fock_L\otimes L^2(\mathcal{Q}_L, d\mu_{\Lambda_L})$.
We use the following identification:
\begin{align}
\mathfrak{H}_L=\int_{\mathcal{Q}_L}^{\oplus} \Fock_L d\mu_{\Lambda_L}(\bphi).
\end{align}
Thus, each vector $\Psi\in \mathfrak{H}_L$
is  a $\Fock_L$-valued measurable map on $\mathcal{Q}_L$, i.e.,
$\bphi \mapsto \Psi(\bphi)$.
Now, we define an antiunitary transformation $\vartheta$ from $\mathfrak{H}_L$ onto 
$\mathfrak{H}_R$ by 
\begin{align}
(\vartheta \Psi)(\bphi)= (\xi\Psi)(r^{-1}(\bphi))\ \
 \mbox{a.e. $\bphi\in \mathcal{Q}_R$, $\Psi\in \mathfrak{H}_L$},
\end{align}
where,
for each $\bphi=\{\phi_x\}_{x\in \Lambda_R} \in \mathcal{Q}_R$,
 we define $r^{-1}(\bphi)\in \mathcal{Q}_L$ by $ \big(r^{-1}(\bphi)\big)_x
=\phi_{r^{-1}(x)},\, x\in \Lambda_L
$.

\begin{rem}
{\rm 
For each measurable function $F(\bphi)\  (\bphi\in \mathcal{Q}_L)$ on $\mathcal{Q}_L$, 
$F(r^{-1}(\bphi))\ (\bphi\in \mathcal{Q}_R)$ can be regarded as a
 function on $\mathcal{Q}_R$.
Let $\Psi\in \mathfrak{H}_L$. By (\ref{ExPhi}), we have the following expression:
\begin{align}
\Psi(\bphi)=
\sum_{\mathbf{X}\in \mathscr{S}_R} \Psi_{r(\mathbf{X})}(\bphi)
 f(r(\mathbf{X}))
\ \ \mbox{a.e. $\bphi\in \mathcal{Q}_L$},
\end{align} 
where 
$\Psi_{r(\mathbf{X})}(\bphi)=\la f(r(\mathbf{X}))|\Psi(\bphi)\ra
$.
 Using this, we have 
\begin{align}
(\vartheta \Psi)(\bphi)=\sum_{\mathbf{X}\in \mathscr{S}_R}
\overline{\Psi_{r(\mathbf{X})}(r^{-1}(\bphi))} e(\mathbf{X})\ \ 
 \mbox{a.e. $\bphi\in \mathcal{Q}_R$}. \ \ \ \ \diamondsuit
\end{align} 
}
\end{rem}

\begin{Prop}
$\vartheta$ satisfies all properties in  (\ref{theta}) and (\ref{theta2}).
\end{Prop}
{\it Proof.} By Lemma \ref{Xi}, it is easy to check that 
$\vartheta a_{r(X)}\vartheta^{-1}=c_X$. 

Note  that the action of the multiplication operator $\phi_x$ is as
follows:
For each $\Psi\in \mathfrak{H}_L$ and $x\in \Lambda_L$, 
\begin{align}
(\phi_x \Psi)(\bphi)= \phi_x \Psi(\bphi) \ \ \mbox{a.e. $\bphi\in
 \mathcal{Q}_L$}.
\end{align}
Thus, we have, for each $\Psi\in \mathfrak{H}_L$ and $x\in \Lambda_R$,
\begin{align}
(\vartheta \phi_{r(x)}\Psi)(\bphi)=\phi_x \xi
\Psi(r^{-1}(\bphi)) 
=(\phi_x\vartheta \Psi)(\bphi)\ \ \mbox{a.e. $\bphi \in \mathcal{Q}_R$},
\end{align}
which implies $\vartheta \phi_{r(x)} \vartheta^{-1}=\phi_x$.

Next, we will prove that $\vartheta \pi_{r(x)} \vartheta^{-1}=-\pi_x$.
Since $\pi_x=-i \frac{\partial}{\partial \phi_x}$, we have, for each
$\Psi\in \mathfrak{H}_L$ and $x\in \Lambda_L$,
\begin{align}
(\pi_x \Psi)(\bphi)=
(-i)\frac{\partial \Psi}{\partial \phi_x}(\bphi) \ \
 \mbox{a.e. $\bphi\in \mathcal{Q}_L$}.
\end{align}
Hence, we have, for each $\Psi\in \mathfrak{H}_L$ and $x\in \Lambda_L$,
\begin{align}
(\vartheta \pi_{x} \Psi)(\bphi)=(+i) \overline{\frac{\partial \Psi}{\partial
 \phi_x}} (r^{-1}(\bphi)) 
=(+i) \frac{\partial (\vartheta \Psi)}{\partial
 \phi_{r^{-1}(x)}} (\bphi)
=-(\pi_{r^{-1}(x)} \vartheta \Psi)(\bphi)
\end{align}
for a.e. $\bphi\in \mathcal{Q}_R$.
Here, we used the fact that $\xi$ is antilinear.
Thus, we conclude that 
$\vartheta \pi_{x} \vartheta^{-1}=-\pi_{r^{-1}(x)}$ for each $x\in
\Lambda_L$, which implies $\vartheta \pi_{r(x)} \vartheta^{-1}=-\pi_{x}$ for each $x\in
\Lambda_R$.
$\Box$

\end{document}